# A computational model of the bottlenose dolphin sonar: Feature-extracting method

Tengiz V. Zorikov

*Institute of Cybernetics, 5 S. Euli, 0186, Tbilisi, Georgia, zorikov@cybernet.ge*

**Abstract.** The data describing a process of echo-image formation in bottlenose dolphin sonar perception were accumulated in our experimental explorations. These data were formalized mathematically and used in the computational model, comparative testing of which in echo-discrimination tasks revealed no less capabilities then those of bottlenose dolphins.

Key words: bottlenose dolphin, echolocation, echo-processing, bio-sonar modeling.

## 1. Introduction

The main goal with which about 30 years ago we have begun exploration of the sonar system of bottlenose dolphins was creation of its technical analogue for the effective solution of the well known real-world problems in the sea. Thereby our attention was focused mostly on the mechanisms of echo-image formation in bottlenose dolphin sonar perception, – the problem of crucial significance in bio-sonar modeling. Our approach to the problem was rather technical then biological, i.e. we were not interested in the constructive peculiarities of a dolphin's sonar analyzer, considering a sonar system of that animal as a "black box". Therefore, saying in advance, the computational model that we have created on the gained data is not an attempt of constructive duplication of bottlenose dolphin sonar or its separate blocks but is its general functional analog.

Intermittently, we were conducting experiments with bottlenose dolphins at the Sevastopol Naval Base, – in total more then 20 experiments, in which dolphins (six different ones in total) were trained to differentiate in passive mode echo-like impulses synthesized by shock excitation of spherical piezoceramic transducer by diverse electrical rectangular pulses, and echoes reflected from different targets.

We used a simple way for mathematical formalization of our experimental findings, just to verify the level of efficiency of the feature-extracting method discovered in our tests.

Considerable part of the data, offered in this paper, was presented at the $151^{st}$ Annual Meeting of the Acoustical Society of America [15].

## 2. The data on echo-processing mechanisms revealed in our experiments

<u>In the experiments with synthetic echoes</u>, we took advantage of the opportunity to make precise monitoring of dolphins' specific reactions to any of physical components combined in these stimuli. It was reached due to a possibility of creation of any desirable composition of components with any combinations of values of these components in simulated signals, – practically unsatisfiable task for a case of actual echoes.

First we re-estimated the size of temporal integration window or the Critical Interval of Time (*CIT*) of the bottlenose dolphins' sonar, – the maximal time interval, within which echo-highlights are still transformed into a merged echo-image in animal perception of echolocation signals. We used two quite different methods and have obtained in both cases about 200 μs, less then 265 μs declared in [1, 6, and 10].

To study the functional role of echo's different physical components in the process of echo-image formation within the *CIT*, we developed a series of logically interrelated tests (16 in total), which were conducted with bottlenose dolphins. The following results were revealed by instrumentality of these tests [2-4, 11-14].

It was shown that dolphins cannot assess echoes' polarity. We can assume thereby that the power spectrum density (*PSD*) of an echo is the basic source of information determining echo-descriptive features (or dimensions) of bottlenose dolphins' sonar perception. The subsequent experiments revealed a string of three hierarchically organized, independent descriptive features, which are defined by different scales of spectral density oscillation of an echo and by its power. Namely, the first, dominant feature in hierarchy depends on coarse-scale oscillations of echo's *PSD* with the periods exceeding ~10 kHz frequency bandwidth (feature *MaPS* – **Ma**crostructure of a **P**ower **S**pectrum). The second or middle in the triad is a feature *MiPS* (**Mi**crostructure of a **P**ower **S**pectrum) being defined by small-scale oscillations of echo's *PSD* with the periods of the interval ~5-13 kHz. (*The partial overlapping of the domains of definition of the MaPS and MiPS features within the interval ~10-13 kHz is a natural thing for bio-systems.*) The last in hierarchy, the minor feature *P* (**P**ower) depends on the echo's overall energy within the *CIT*.

The hierarchy between the descriptive features became apparent in the process of differentiation of stimuli by dolphins. In order to distinguish signals, dolphins compared successively their features' values from senior to minor, terminating the process at the feature, which contained detectable differences in presented stimuli (the distinctive feature). I.e. the features of less meaning in hierarchy then the distinctive one were always disregarded by dolphins.

The rigidity of hierarchy between ascertained features has been confirmed also during subsequent identification by dolphins of the signals having been stated in the previous signal-discrimination tests as the reference stimuli, or the stimuli, dolphins' swim responses to which were accompanied by food reinforcement. Analyzing for that purpose probe signals, dolphins verified not only coincidence in the values of the feature being chosen by animals as the distinctive in the previous task but, first of all, coincidence in the values of features of higher position in hierarchy then the distinctive one. Otherwise such probe signals were rejected by animals. I.e. the values of features of higher meaning then the distinctive one have been yet verifying by dolphins during the process of identification of the reference stimuli despite of their uselessness in the previous signal-discrimination tasks because of equality of their values in compared stimuli.

All in all, the rules describing the procedures of echoes discrimination and their identification in bottlenose dolphins can be formulated as follows. – A bottlenose dolphin, when distinguishing signals, compares successively their descriptive features' values from senior to minor, terminating the process at the feature, which contains detectable differences in compared stimuli (the distinctive feature). Herewith if a dolphin utilizes some feature as the distinctive one in a previous signal-discrimination task, then in order to preserve the image of the reference signal in the animal's perception, it is necessary and sufficient to preserve the same values of the distinctive feature and all higher ones in order of hierarchy.

Fig. 1 gives graphical interpretation of the procedures described above.

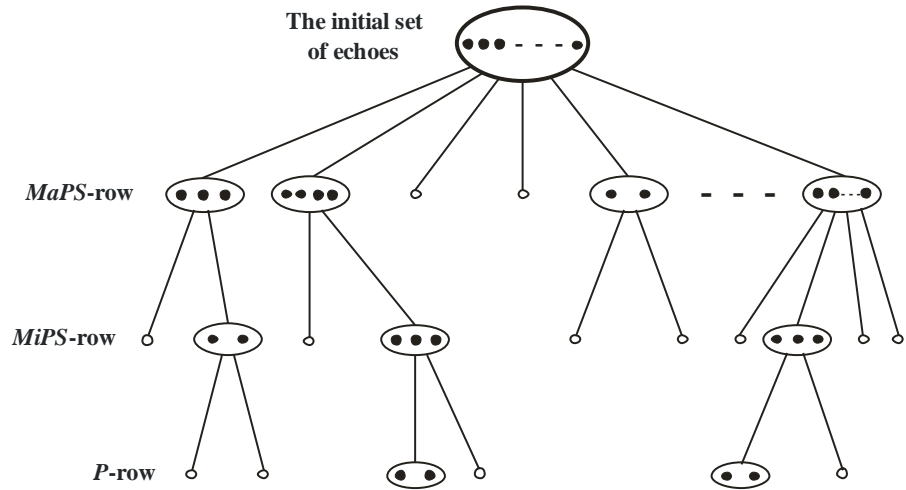

Fig. 1. The initial set of echoes (filled circles in the top ellipse) is parted on the first step in accordance with their *MaPS* values. Obtained subsets can contain one or more echoes (single circles or ellipses of the *MaPS*-row). The process of differentiation is ended for echoes of one-element subsets (circles). The identifiers for those echoes will be the *MaPS* values of appropriate nodes of the *MaPS*-row. The multielement subsets of that row (ellipses) are parted on the second step by comparison of echoes' *MiPS* values (independently for each subset). The subsets obtained as a result are represented by the *MiPS*-row of the graph. As earlier, differentiation is ended for echoes of one-element subsets; however the identifiers for separated echoes will be the pairs of *MaPS*-*MiPS* values of the nodes of trajectories appropriate to those echoes. On the third step, each of the multielement subsets of the *MiPS*-row are parted independently by comparison of *P* values of echoes. The identifiers for echoes of one-element subsets of *P*-row will be the triads of *MaPS*-*MiPS*-*P* values of the nodes of trajectories appropriate to those echoes. The echoes grouped in multielement subsets of *P*-row (ellipses) will be attributed as the indistinguishable/coincident ones.

In the experiments with actual echoes, dolphins were trained to discriminate echoes from actual targets beforehand recorded onto magnetic tape (within 5-200 kHz bandwidth), separately off animals [11]. The test targets were sounded by signals simulating bottlenose dolphin's on-axis emitted echolocation clicks, with repetition rate of 20-80 Hz.

We used ultrasonic tape-recorder and passive mode in echo-discrimination tests with dolphins because by such way we could conduct thorough comparative analysis of echoes being distinguished and inaccessible to be distinguished by animals and make well-defined conclusions without any doubts caused by a number of factors, hardly traceable otherwise.

Registration of test echoes reflected from various types of mines and their dummies, as well as from moving skin-diver in four different aspect angles was fulfilled at a distance of 10-15 m from targets in natural noise conditions of a bay coastal zone that have provoked huge spectral fluctuations in recorded signals, especially in the echoes from moving skin-diver. Despite of that, majority of trains of those echoes were distinguished by dolphins with about 90-100% of correct responses.

This suggested the idea on the mechanism of averaging by dolphins the features values over a series of echoes. – Appropriate comparative calculations have confirmed the validity of that hypothesis. We assume this is rather significant mechanism that provides bottlenose dolphins with exclusively high level of noise immunity.

## 3. The model of echo-processing

We used a simple way for mathematical formalization of the represented above data just to verify the level of efficiency of the feature-extracting method discovered in our experiments.

The first in hierarchy, senior feature *MaPS*. – In accordance with above stated definition, the feature *MaPS* must be described by the function, sensitive towards the coarse-scale oscillations of echo's *PSD* (with the periods exceeding 10 kHz) and invariant relatively smaller-scale oscillations of echo's *PSD* and its energy. The role of such function fulfills a 16-dimensional vector, which components are calculated by integration of the echo's normalized *PSD* within 10 kHz sub-intervals across the range of dolphin's sonar frequency perception (30-190 kHz); in total 16 subintervals (components) covering this range (1):

$$mac_j \bigg|_{j=\overline{1,16}} = \frac{\int_{f_{j-1}}^{f_j} |F\{x(t)\}|^2 df}{\int_{f_0}^{f_{16}} |F\{x(t)\}|^2 df} \quad (1)$$

where:
  $F$ – an operator of Forward Fourier Transformation;
  $f_0$ = 30 kHz – the bottom boundary of sonar frequency perception;
  $f_{16}$ = 190 kHz – the upper boundary of sonar frequency perception;
  $f_j - f_{j-1}$ = 10 kHz – intervals of integration.

The second in hierarchy feature *MiPS*. – The small-scale oscillations of echo's *PSD* within the frequency range of 5-13 kHz (see definition given above) correspond in temporal expression to echo's cepstral shape in the range of ~75-200 µs (the reciprocals of the frequency boundaries). Therefore, for calculation of the value of feature *MiPS*, we integrate the echo's normalized cepstral function within 5% sub-intervals (in accordance with bottlenose dolphin's difference liman for time delays [9]) across the range of 75-200 µs. As a result, we obtain 19-dimensional vector, the components of which are calculated by the formula 2:

$$mic_j \bigg|_{j=\overline{1,19}} = \frac{1}{[1+0.05(j-1)]} \int_{\tau_{j-1}}^{\tau_j} \left| F^{-1}\left\{ \ln\left[ \frac{|F\{x(t)\}|^2}{\int_{f_0}^{f_{16}} |F\{x(t)\}|^2 df} + 1 \right] \right\} \right|^2 d\tau \quad (2)$$

where:
  $F^{-1}$ – an operator of Inverse Fourier Transformation;
  $\tau_0$ = 75 µs – the bottom boundary of feature *MiPS* domain of definition;
  $\tau_{19}$ = 200 µs – the upper boundary of feature *MiPS* domain of definition;
  $\tau_j - \tau_{j-1} = 0.05 \tau_{j-1}$ – intervals of integration.

Fig. 2 gives an example of two signals and their *MaPS* and *MiPS* histogram images calculated in accordance with the formulas 1, 2.

The last in hierarchy, the minor feature *P* we calculate simply integrating echo's *PSD* over the diapason of sonar frequency perception 30-190 kHz (3):

$$P = \int_{f_0}^{f_{16}} |F\{x(t)\}|^2 df \qquad (3)$$

The feature-averaging mechanism is applied in our model as follows. We assume that each target to be recognized is represented by a sufficiently great number of echoes registered in adequate natural conditions. The standard echolocation click ought to be utilized for sounding of the targets. First, the average values for all features (*MaPS*, *MiPS* and *P*) are calculated on a sufficiently great number *N* of echoes, – the feature standards of the given target. The same operation is carried out on a sufficiently great number *M* of various subsets of the target echoes, each of which includes *n* echoes. Thus we obtain *M* average values for each feature (*MaPS, MiPS* and *P*), – the training averages. Next, the distances between features' standards and appropriate training averages are calculated (*M* distances for each feature) and the confidence intervals determining the range of dispersion of the training averages relatively the appropriate standards are calculated for each feature. Finally the set of these six parameters (three feature standards and three confidence intervals appropriate to them) are stored in model's database as the image of the given target. Such easy way of accumulation of the targets data (independently, i.e. without any mathematical operations between targets' data themselves) allows updating the database by simple adding to it any new images (the sets of six parameters mentioned above).

The values of *N*, *M* and *n* determine the level of noise immunity and accuracy of the model. It is clear that in a definite practical task these parameters must be determined experimentally, in adequate natural conditions. Currently, these parameters are changeable in the model, and we used their diverse values in the measurements presented below, depending on the complexity of particular tasks.

The process of identification of targets in the model is running similarly to that of dolphins. The average value for each feature (*MaPS*, *MiPS* and *P*) is calculated over a set of *n* echoes (the same *n* as in the training subsets) from a target liable to identification (the target have to be sounded by the standard echolocation clicks), – three probe averages. These averages are comparing one after another (from senior to minor) with the appropriate feature standards stored in the data base. First, the model compares the *MaPS* probe average with each of *MaPS* standards. The following variants are possible:

1. The *MaPS* probe average does not meet the limits of *MaPS* confidence intervals stored in the database. – The target is admitted as unknown;

2. The *MaPS* probe average meets the confidence interval of a certain *MaPS* standard stored in the database. – The identification is made. The name of a target having this *MaPS* standard is attributed to interrogated target;

3. The *MaPS* probe average meets the confidence intervals of *MaPS* standards of the subset of targets stored in the database. – The process of recognition is continuing within this subset of targets, although the *MaPS* feature is replaced by the second in hierarchy feature *MiPS*.

The possible outcomes of the second step of recognition and their interpretation are the same as above. In the case of the second variant, identification comes to an end, and the name of a certain target coinciding now with the interrogated one in the values of features

*MaPS* and *MiPS* simultaneously is attributed to the interrogated target. As well as earlier, the process of identification is continuing only in the third case, when the *MiPS* probe average meets the confidence intervals of the *MiPS* standards of several targets from the subset grouped on the first step of identification. The process of recognition is continuing now within this new subset of targets (coinciding with interrogated one in the values of features *MaPS* and *MiPS* simultaneously), and the last step of comparison is carried out now in the values of the feature *P*. Again, the model uses the same logic for the outcomes of the third step. In case of success (the second variant) interrogated target is considered as identified, but already with the help of all three features now.

## 4. Performance of the model

To study the model's performance, we used two-highlight synthetic echoes simulated in a computer in accordance with the formulas:

$$x(t) = f(t) + af(t-d), \quad \text{where} \quad f(t) = \begin{cases} 0, & \text{if } t < 0 \\ e^{-0,1t} \cdot \sin 0.875t, & \text{if } t \geq 0 \end{cases} \quad (4)$$

By this way, we simulated signals similar to those, which were used in the experiments with bottlenose dolphins [9]. However we have additionally complicated the discrimination tasks for the model, randomizing the time delay of the second highlight (the coefficient *d* could be varied randomly within desirable range under the bell-shaped distribution of probabilities) and mixing signals with white noise of definite intensity.

Difference liman of the model (*DLT*) for time delays (*ΔT*) of the second highlights was assessed in the first measurement. The coefficient *a* was equal to 1. Time delays (*d*) of the second highlights in compared signals were varied randomly around the center *ΔT* values within the intervals [*ΔT*–0.2*ΔT*, *ΔT*+0.2*ΔT*]. In conducted measurements, the share of intersection of the variability intervals in discriminated signals was changed from 50% up to about 95 % along with the *ΔT* increasing (Fig. 2).

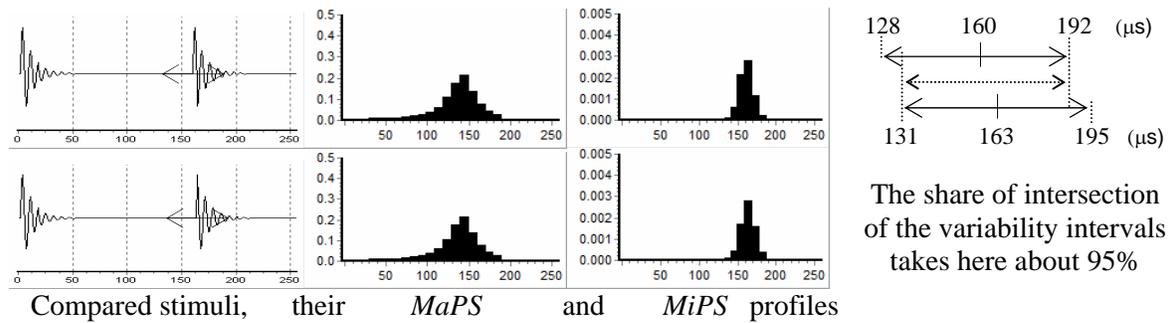

Compared stimuli, their *MaPS* and *MiPS* profiles

Fig. 2. Example of two signals having been discriminated by the model (left column) and their *MaPS* and *MiPS* features' averaged histogram images (next to them). The center delays of these signals *ΔT* are equal to 160 and 163 μs, i.e. the *DLT* of the model in this case is 3 μs (filled circle in Fig. 3). The limits of random variation of delays for the top signal are 128-192 μs and 131-195 μs for the bottom one (right plot). Thus the share of intersection of the intervals of variation takes here about 95%.

Parameters *N* and *n*, determining the sensitivity of the model, were installed accordingly to a complexity of a task. In the given case the number of echoes averaged in the feature standard calculation (*N*) was equal to 100 ones, the number of echoes utilized for training and probe average calculation (*n*) was equal to 50 ones. Despite of random distribution of the delays in our measurement, the results (Fig. 3, circles) look better, then those of dolphins [9].

We used the same synthetic echoes in the next measurement, adding to them white noise. The signal-to-noise ratio was equal to 12 dB. Here we increased *N* up to 200 echoes, and *n* – up to 100 ones. Adding noise has caused deterioration of the thresholds for $\Delta T > 40$ μs (Fig. 3, triangles) in comparison with the previous data (circles). Nevertheless, conducted measurements yet demonstrate definite advantage of the model in discrimination of time intervals. It is presumable as well to expect better sensitivity of the model within the area 0-40 μs, limited here by 1 μs. The point is that it could be caused by the rate of digitization, which is equal to 1 μs in our system, and we simply could not check up the values of less meaning. In any case, obtained data are already better then those of dolphins, taking additionally into account the noise and stochastic character of distribution of delays.

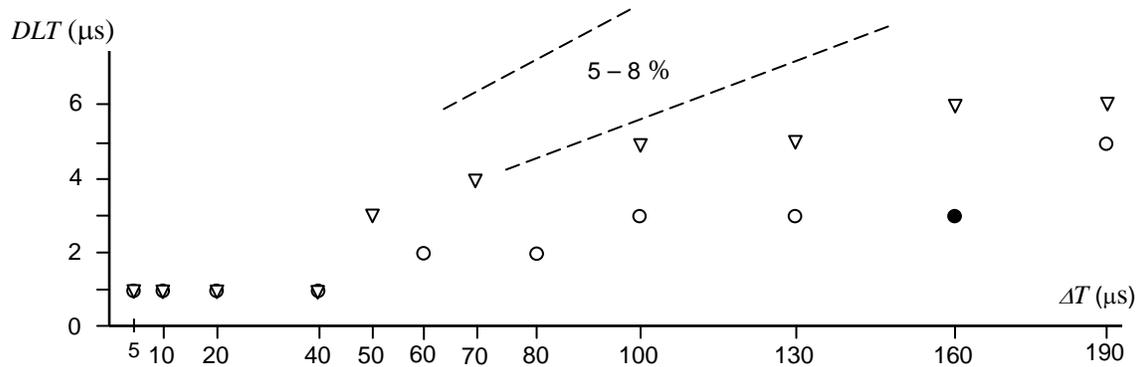

Fig. 3. Difference liman of the model for time delays of the second highlights (*DLT*) for pure signals (circles) and signals mixed with white noise (triangles). The area limited by dashed lines corresponds to the *DLT* (5-8 %) revealed in experiments with dolphins [9]. Filled circle marks a threshold for the case given as the example in Fig. 2.

Under the definition given above, feature *MiPS* must "start working" from 75 μs; however, in conducted above measurements this feature "took up initiative" from $\Delta T > 50$ μs. It can be explained by penetration of the information into the definitional domain of the *MiPS* feature (75-200 μs) because of random distribution of the starting positions of second highlights and definite duration of these highlights themselves.

In the next measurements, we utilized the data represented in [5]. Two-highlight synthetic echoes, having differences in the center frequencies of the second, low-amplitude highlights, were used in that work for investigation of the mechanisms of echo-processing in bottlenose dolphins. The energy ratio between the first and second highlights (*ΔdB*) and the timing relationship (*ΔT*) between the first and second highlights were manipulated. An iso-sensitivity function, demonstrating the ability of animal to discriminate such type of stimuli, was derived (Fig. 5, stars).

To compare the model's performance with the data obtained in above mentioned work, we used signals, in which low-amplitude second highlights were synthesized by summation of the second and third components of the formula 5, distanced alternatively by 7 or 10 μs relatively each other:

$$x(t) = f(t) + af(t - d_1) + af(t - d_2) \tag{5}$$

where the time delay between second and third components $(d_2 - d_1)$ was equal to 7 or 10 μs in discriminated signals. Resulted second highlights were about 40 μs in duration and differed in their waveforms and *PSD* (Fig. 4).

The coefficient *a* was manipulated within the interval [0.003, 0.06]. Measurements were conducted on a background of white noise of –40 dB relatively the first highlight, as well as it was made in experiments with a dolphin [5]; parameter *N* was equal to 100 units, parameter *n* – to 50 ones.

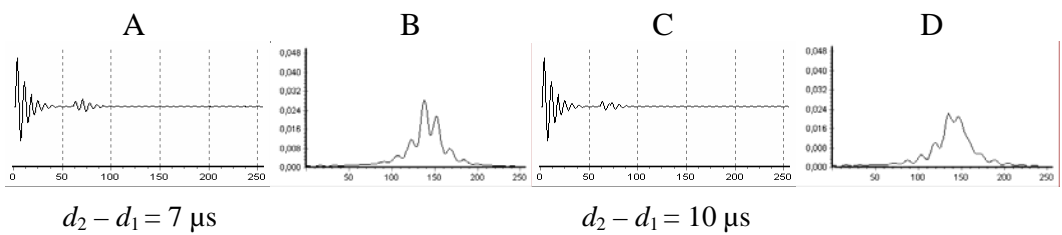

$d_2 - d_1 = 7$ μs    $d_2 - d_1 = 10$ μs

Fig. 4. Example of two-highlight synthetic echoes (A, C) and their *PSD* ((B, D), simulated in a computer. The time delay *ΔT* of the second highlights is equal to 60 μs in the given case. To make visible the differences in *PSD* of compared signals and in waveforms of the second highlights in demonstrated pictures, we excluded noise and made the coefficient *a* equal to 0. 1.

Obtained in this study results demonstrate advantage of the model over the dolphin (Fig. 5, stars for the dolphin's performance, circles and triangles for the model's performance).

It was suggested in the paper above mentioned that dolphins processed multiple echo highlights as separable analyzable features in the discrimination task, perhaps perceived through differences in spectral rippling across the duration of the echoes.

In our case the model operates both of features *MaPS* and *MiPS*, which alternately show excellent performance (-40 dB and even less within *ΔT* = 0-5 μs, ellipse in Fig. 4), each in its own definitional domain. The *MaPS* thresholds increased at *ΔT* = 70 μs, remaining roughly constant (between -30 and -25 dB) up to 190 μs (Fig. 4, circles). It is happening because the difference in coarse scale structure of *PSD* within the range of 100-190 μs (i.e. outside the definitional domain of feature *MaPS*) is contributed only by a difference in the waveforms of low amplitude secondary highlights (due to equality in waveforms of the first highlights). Therefore, the difference in waveforms of the secondary highlights remains invariable along this range.

In turn, the feature *MiPS* provided perfect distinction of signals (-40 dB) in the range of 70-190 μs (Fig. 4, triangles). Difference in the *MiPS* values of compared signals within this interval is provided by the difference in delays of their third components relative to the first one (formula 3). This difference is equal to 3 μs, and remains constant within all the range of *ΔT* variation. This explains the constancy of the *MiPS* thresholds of the model across the range of definition of this feature.

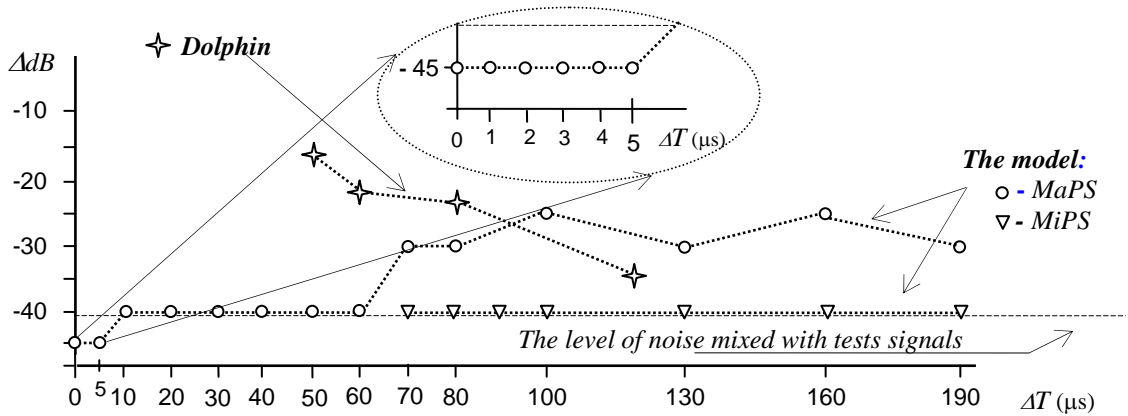

Fig. 5. Iso-sensitivity function demonstrating the model's performance in discrimination of signals with low-amplitude second highlights (circles for the *MaPS* feature and triangles for the *MiPS*). Stars demonstrate performance revealed in the experiments with dolphins [5].

We tested the model on echoes from actual targets in collaboration with Dr. Patrick W. Moore (SPAWARSYSCEN, San Diego, CA, USA). We used echoes that were captured and digitized at 500 kHz during a matching-to-sample task performed by dolphin [7, 8]. The targets utilized for the matching tasks were fuel bottles containing different fluids. All targets were identical in their outer form, differing only in their internal constituents.

In our measurements, the model distinguished echoes reflected from bottles filled by fresh water, salty water, glycerol, kerosene, and air. Confident identification of echoes from the bottles filled by air, kerosene, and glycerol was reached by the model when the number of echoes forming the features standards ($N$) was increased up to 10 units, and the number of echoes forming the probe and training averages ($n$) – up to 5 units. The identification of echoes from the bottles filled by water and salty water had required increasing these parameters: $N$ – up to 25 units, and $n$ – up to 15 ones. Herewith the model used only the feature *MaPS* in all cases for solving the tasks.

## 5. Conclusion

To study the mechanisms of echo-image formation, we used in signal-discrimination tests with bottlenose dolphins echo-like simulated stimuli. By instrumentality of this, in the series of logically interrelated experiments (16 in total), we have managed to discover the total set of independent descriptive features (dimensions) forming the echo's image in dolphin's sonar perception, have proved existence of hierarchy between these features, and have displayed the structure of this hierarchy. The detailed description of the procedure of echo-discrimination and the decision rule describing the process of echo-identification in bottlenose dolphins have been established in these experiments as well.

However we could miss some of the specific echo-processing mechanisms, using in the tests echo-like impulses only. This lack was balanced somehow in the experiments with actual echoes registered beforehand in a tape-recorder and having been presented to animals afterwards from the tape-recorder. By such way, we could conduct comprehensive comparative analysis of the test echoes and revealed as a result the mechanism of statistical processing of echo-series by dolphins. Thus, in aggregate with the data of the experiments with echo-like stimuli, we have obtained practically all qualitative and quantitative parameters needed for starting the sonar-modeling process.

The computational model represented in this work is only the first step made just for the purpose of verifying the level of efficiency of the feature-extracting method discovered in our experiments. It is clear that the next testing phase will require more developed approach. Nevertheless, our model already demonstrates even better performance then that of bottlenose dolphins (at least, upon the set of signals utilized in above presented measurements). Herewith we should also have in view that the level of accuracy and noise stability of the model demonstrated above is only a particular effect of the values of parameters determining this level in the model and installed definitely for the given cases. We could reach more efficient results, having altered these parameters properly.

As well important is that such high level of efficiency of the model is ensured by very simple and time-saving computational algorithms providing the real-time echolocation, – the necessary requirement to any sonar system.

**References**

[1] Au, W.W.L., Moore, P.W.B., and Pawloski, D.A., (1988), "Detection of complex echoes in noise by an echolocating dolphin", J. Acoust. Soc. Am. 83, 662-668.
[2] Dubrovskiy, N.A., Zorikov, T.V., Kvizhinadze, O.Sh., Kuratashvili, M.M., (1991), "Feature description of signals and principles of its organization in the auditory system of bottlenose dolphin", Akusticheskii zhurnal, Vol. 37, No. 5, 933-937.
[3] Dubrovskiy, N.A., Zorikov, T.V., Kvizhinadze, O.Sh., Kuratashvili, M.M., (1992), "Feature description of acoustic signals under their discrimination by bottlenose dolphin", Sensornye sistemy, Vol. 6, 37-48.
[4] Dubrovskiy, N.A., Zorikov, T.V., Kvizhinadze, O.Sh., Kuratashvili, M.M., (1992), "Feature description of signals and principles of its organization in the auditory system of bottlenose dolphin", American Institute of Physics. Physical Acoustics, 485-487.
[5] Helweg, D.A., Moore, P.W., Danciewicz, L.A., Zafran, J.M., Brill, R.L., (2003), "Discrimination of complex synthetic echoes by an echolocating bottlenose dolphin", J. Acoust. Soc. Am., Vol. 113, No. 2, 1138-1144.
[6] Moore, P.W.B., Hall, R.W., Friedl, W.A., and Nachtigall, P.E., (1984), "The critical interval in dolphin echolocation: What is it?", J. Acoust. Soc. Am. 76, 314-317
[7] Roitblat, H.L., Penner, R.H., & Nachtigall, P.E. (1990) "Matching-to-sample by echolocating dolphin", J. of Experimental Psychology: Animal Behaviour Processes, 16, 85-95.
[8] Rootlet, H.L., Moore, P.W.B., Helweg, D. A. and Nachtigall P.E., (1993), "Representation and Processing of Acoustic Information in a Biomimetic Neural Network", Animals to Animals 2: Simulation of Adaptive Behavior", J.-A Myer, H.L. Roitblat, & S.W. Wilson (Eds.), MIT press, 1-10.
[9] Vel'min, V.A., and Dubrovskiy, N.A., (1975), "On the auditory analysis of pulse sounds", Dokladi Akademii Nauk SSSR, 225, No. 2, 229-232.
[10] Vel'min, V.A., and Dubrovskiy, N.A., (1976), "The critical interval of active hearing in dolphins", Sov. Phys. Acoust. 2, 351-352.
[11] Zorikov, T.V., (1985), "Feature description of signals and principles of its organization in the auditory system of bottlenose dolphin", Candidate dissertation, St. Petersburg (former Leningrad), 137.
[12] Zorikov T.V., Dubrovskiy N.A., Beckauri N.J., (2001), "Signal processing by the Bottlenose dolphin's sonar: experiments and modeling", 2nd Symposium on underwater bio-sonar and bioacoustics systems, Vol.23, Pt.4, Loughborough, UK, 65-74.

[13] Zorikov T.V., Dubrovskiy N.A, (2003), "Echo-Processing Procedure in Bottlenose Dolphins", OCEANS'2003 Conference, San Diego, USA, # 656.
[14] Zorikov, T.V., Moore, P.W.B., Beckauri, N.J., (2004), "A model of echo-processing in bottlenose dolphins", Symposium on Bio-Sonar Systems and Bio-Acoustics, Vol.26, Pt.6, Loughborough, UK, 73-81.
[15] Zorikov T.V., (2006), "A computational model of bottlenose dolphin sonar: Feature-extracting method", J. Acoust. Soc. Am., Vol. 119, No. 5, 3317. [Invited presentation]